\documentclass[journal=cmatex,manuscript=article]{achemso}
\usepackage[version=3]{mhchem}
\usepackage{amsmath}
\usepackage{amsfonts}
\usepackage{braket}
\usepackage{array}
\usepackage{graphicx}
\usepackage{caption}
\usepackage{subcaption}
\usepackage{color}
\usepackage{booktabs}
\usepackage{threeparttable}
\usepackage[dvipsnames]{xcolor}
\usepackage[normalem]{ulem}

\usepackage{overpic}

\usepackage{xr}
\usepackage{pdfpages}

\title{A Machine-learning-aided Workflow for Sampling Small Polaron Dynamics in Energy Materials}

\author{Tavinder Singh}
\affiliation{Chair for Theoretical Physics VII and Bavarian Center for Battery Technologies, University of Bayreuth, Universit\"atsstr. 30, 95447 Germany}

\author{Harald Oberhofer}
\email{harald.oberhofer@uni-bayreuth.de}
\affiliation{Chair for Theoretical Physics VII and Bavarian Center for Battery Technologies, University of Bayreuth, Universit\"atsstr. 30, 95447 Germany}
\date{\today}

\begin{document}
\begin{abstract}
One of the defining characteristics of most energy materials, used as e.g.~battery electrodes, photovoltaics, or photocatalysts, is their ability conduct charges such as electrons and holes. The exact mechanism of this conductivity depends on the material, but is often dominated by the existence of mobile small polarons, i.e.~electrons or holes strongly localised through the polarisation of their environment. Computationally, small polarons can be treated with reasonably accuracy through the use of self-interaction corrected or hybrid density functional theory (DFT) together with local geometry optimisation. Similar approaches can be employed to determine the energy barriers to movement of the polarons through hopping. Unfortunately, realistic energy materials tend have very large numbers of symmetry inequivalent polaronic configurations and correspondingly large numbers of hopping barriers, making them hard to tackle explicitly.

In this work we therefore extend the computational reach of DFT calculations with machine-learned (ML) surrogate models, which then feed into a kinetic model to ultimately quantitatively estimate the conductivity of a battery anode material (Li$_4$Ti$_5$O$_12$ or LTO) at realistic, defected conditions. We calculate the conductivity at room temperature to be 6.7 $\times$ 10$^{-3}$ S/cm which lies well in the experimental range. To this end, we only needed a total of 154 polaron barrier calculations, out of 3649 symmetry inequivalent ones, achieving an over 20-fold increase of computational efficiency. Through the use of explainable ML models and input features, we further were able to extract the most important features determining the height of the hopping barrier, to find that the interaction of the polaron with the defects. Overall, our work demonstrates how DFT-trained ML surrogates can be used to quantitatively and efficiently sample the dynamics of small polarons in energy materials.
\end{abstract}

\section{Introduction}
The performance of energy generation and energy storage materials is strongly influenced by the efficient movement of charge carriers in the material \cite{righetto2023cation,kick2020mobile, kick2021polaron}. Mechanistically, such charge transport can occur predominantly in two ways. \cite{oberhofer2017charge, blumberger2015recent} In band transport, charge carriers are delocalized and propagate through extended electronic states, whereas in hopping transport, they can localize through their interaction and subsequent deformation with the lattice in the form of a small polaron quasiparticle.\cite{holstein1959studies_1,holstein1959studies_2} Transport of a small polaron occurs by thermally activated hopping between localized sites \cite{oberhofer2017charge}. 

The occurrence of small polarons can drastically alter the properties of a material. For example, Li\textsubscript{4}Ti\textsubscript{5}O\textsubscript{12} (LTO), which is used as an anode in Lithium-ion batteries (LIBs),\cite{zhang2022li4ti5o12, julien2024fabrication, thackeray2021li4ti5o12} was found to only conduct due to a polaron hopping mechanism \cite{kick2020formation, kick2020mobile} and would, otherwise, be a strong insulator. Therefore, any theoretical treatment of LTO or similar materials needs to address the polaron dynamics or risk being incomplete.

Computationally, the study of polaron localization and transport is generally complicated by the electron delocalization error associated with computationally-cheap semi-local density functionals, all but preventing the occurrence of polaronic states.\cite{deskins2011distribution,shibuya2012systematic}
However, this can be mitigated through the use of hybrid functionals which have been shown to accurately model polarons, albeit at very high computational costs. Moreover, in a simulation cell, there can be thousands of inequivalent polaron and hopping barrier configurations, which can make a proper sampling of polaron space cumbersome. Alternatively, the Hubbard-corrected DFT+U occupation matrix control method has been shown to model polaron localization energies and barriers at a more reasonable cost.\cite{kick2020mobile} Unfortunately, sampling thousands of configurations and barriers is still often beyond the computational capabilities available for the study of a single material.

Instead, recent advancements in the field of machine learning (ML) have shown promising results in predicting various material properties from   data sets.\cite{yao2025adapting, yao2024designing} 
Indeed our earlier work, \cite{10.1021/acs.chemmater.6c00709}
has demonstrated the use of small-data ML models as a filter for energetically relevant polaron configurations, saving ten thousands of core hours in compute time.

In this work, we utilize a similar approach to predict dynamical properties of polarons. Combining electronic density functional theory (DFT), ML, and a kinetic Monte Carlo (kMC) dynamics approach, we are able to calculate the conductivity of LTO to a high degree of accuracy. Given a specific supercell of LTO, we first grouped similar polaron configurations together, resulting in seven distinct clusters. We then randomly selected 161 configurations from these clusters for DFT calculations. With this, we ensured a diverse training dataset, which is crucial for small data learning. We then trained three different ML-models; support vector regression (SVR), gaussian process regression (GPR), and gradient boosted decision trees (GBDT)
on the DFT calculated barriers, and found GBDT to be the outperformer for our regression task. The trained GBDT model was then utilized to predict the hopping barrier for all 3649 configurations. To predict the macroscopic properties, we performed kMC simulations using the trained GBDT model as a barrier calculator for the kMC simulations. The calculated conductivity was found to be in good agreement with the experimental study performed on Ta-doped reduced LTO. We showed that any more approximate calculation of the barriers, such as, e.g.~taking only the mean barrier height of the test set, would result in a much different conductivity. 

\section{Methods}
\subsection{Computational Details}
All first-principles, spin-polarized calculations have been performed using the FHI-aims code.\cite{blum2009ab} The generalized gradient approximated (GGAs) Perdew Burke Ernzerhof (PBE) functional \cite{perdew1996generalized} was used to treat the exchange-correlation energy, while an adequate electron localization has been achieved using the DFT$+$U correction \cite{anisimov1991band, kick2019intricacies}. Specifically, we used the rotationally invariant form of $+$~U and a fully localized limit to treat the double counting error \cite{himmetoglu2014hubbard}. Atomic like basis functions served as Hubbard projectors for Ti 3d, and a $+$~U value of $2.65$ eV was used, similar to previous studies \cite{kick2020formation, kick2020mobile}. Numerical convergence has been achieved using standard \emph{tight} \emph{tier1} basis sets.\cite{kick2020formation,TSingh2026Localization}
However, also following our earlier work we excluded the \texttt{tier-1 3d hydro} basis functions, as they have been found to interfere with the $+$~U correction.\cite{kick2020formation, kick2020mobile} Overall, these numeric basis settings are roughly equivalent to a polarized triple-$\zeta$ split valence Gaussian type orbital basis set \cite{lamiel2017anatase}. For reciprocal space sampling, we used a 2$\times$2$\times$1 $\Gamma$-centered k-grid sampling.  Polarons were formed from the two unpaired electrons resulting from an oxygen vacancy, which we created in a 2$\times$2$\times$1 (Li$_{32}$Ti$_{40}$O$_{96}$) LTO supercell. Then, to localize and migrate the polaron, we used the $+$U occupation matrix control method.\cite{kick2020mobile} Polaron migration barriers were approximated at the mid point between the initial and final state. 

\subsection{Input features}
Previous studies suggest that localization and migration are influenced by polaron$-$polaron and polaron-defect interactions.\cite{kick2020mobile, kick2020formation} In our earlier work \cite{10.1021/acs.chemmater.6c00709},
we found that polaron localization energies can be well described with a machine-learned model resting on a set of purely structural features based on distances between polarons, the polarons and the defect, and polarons and Li ions. Based on these findings, we selected similar structural features for the description of hopping barriers. All of the input features can be extracted from a ground state structural model, without any costly computations. In detail, we include features that can describe the interaction between the migrating polaron -- both at the initial and final positions -- and the other static polaron. In the same way, we also consider the interaction between both polarons and the defect. 

\begin{figure}
    \centering
    \includegraphics[scale=1.0]{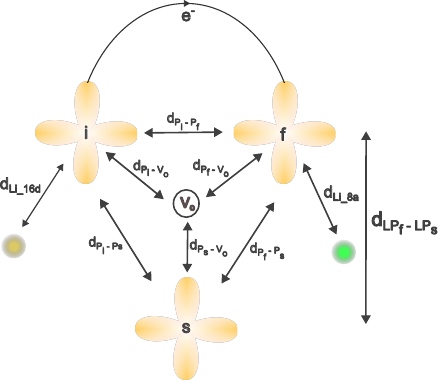}
    \caption{Depiction of distance features for LTO 2$\times$2$\times$1 supercell. The orbital illustration represent the initial, final and static polaron, with the encircled \emph{V$_o$} represents the oxygen vacancy, and the yellow and green coloured balls represent the Li-16d and Li-8a ions.}
    \label{fig:figure_1_features}
\end{figure}

We employ the following nomenclature: the subscript \emph{s} represents a static polaron, while the subscripts \emph{i} and \emph{f} represent the mobile polaron at its initial and final positions, respectively. The first set of important features can then be represented as $d_{P_x-P_y}$, where $x$ and $y$ represent any polaron pair, and $d_{P_x-V_o}$, represents the distance of any polaron from the oxygen vacancy. To account for Ti-layers, we considered vertical distances between polaron sites; represented as $d_{LP_x-LP_y}$. To describe the interaction with lithium (Li) ions, we considered different features based on their Wyckoff sites, 8a and 16d due to fractional occupation of Li/Ti at the 16d site.\cite{kick2020mobile} Thus, $d_{P_x -\text{Li-8a}}$ and $d_{P_x -\text{Li-16d}}$ represent the distances from the nearest 8a and 16d Li-atom, respectively. A schematic of these structural features considered for barrier learning purpose is shown in Fig.~\ref{fig:figure_1_features}

To account for the mid-point of the transport pathway, we considered a ghost atom at the middle of the initial and final sites, and all the features with distance from the ghost atom contain a subscript $Gh$. Furthermore, the likely features were further combined to form sets, in order to minimize the complexity and for symmetrization purposes. The features were symmetrized using radial Behler-Parinello symmetry functions, described by the following equation:\cite{PhysRevLett.98.146401} 
\begin{align}
    G = \sum_{j=1}^{N} e^{-\eta (d_{ij}-R_s)^2}
\end{align}
where, N = 3 represents the number of polaron positions considered here: static, initial and final, $d_{ij}$ is any of the distance features of interest, $\eta$ a width parameter and $R_s$ is a shift parameter. For our study, we set \(\eta = 0.1\) and \(R_s = 3.0\) Å based on the values used in our previous study.\cite{10.1021/acs.chemmater.6c00709}
A full description of the structural features is given in the supplementary file, Table S1, and after symmetrization, the final feature used to train the ML model is listed in Table S2. In total, we used nine symmetrized structural features to train the ML model for the polaron migration barrier. 

\subsection{Barrier-Configuration Classification}
The amount of data required for training an ML-model depends on the complexity of the target, independent features, and the selected ML-algorithm.\cite{xu2023small} Due to the high computational cost of every single data point, i.e.~for each polaron migration barrier, we would ideally like to generate only a minimum of training data points. Unfortunately, such low data learning can easily lead to over or under fitting.\cite{xu2023small} 
Moreover, not only for low data learning approaches, it is beneficial to make the training data samples as diverse as possible, uniquely sampling the configuration space, so that the training set covers all physically distinct hopping environments.\cite{10.1021/acs.chemmater.6c00709} There are a number of methods that can be employed to perform such a task, such as partition based methods that include K-means\cite{orlando2021investigation}, and K-medoids, and hierarchical clustering-based methods that include agglomerative (bottom-up) and divisive (top-down) clustering.\cite{bishop2006pattern, james2013introduction}. For our study, we classify the entire sample space of barrier configurations using agglomerative clustering with the ward linkage criterion implemented in the \emph{scipy} python library.\cite{virtanen2020scipy} Agglomerative clustering assumes that each sample is an independent cluster in the beginning. With a gradual increase in the linkage distance, these independent clusters fuse to form a larger cluster. At smaller linkage distances, configurations in the clusters are generally more similar to each other, while at higher linkage distances, clusters also contain more dissimilar configurations. In our case, clustering is based on the set of input features, i.e.,~distances, discussed above.

\begin{figure}
    \centering
    \includegraphics[scale=0.8]{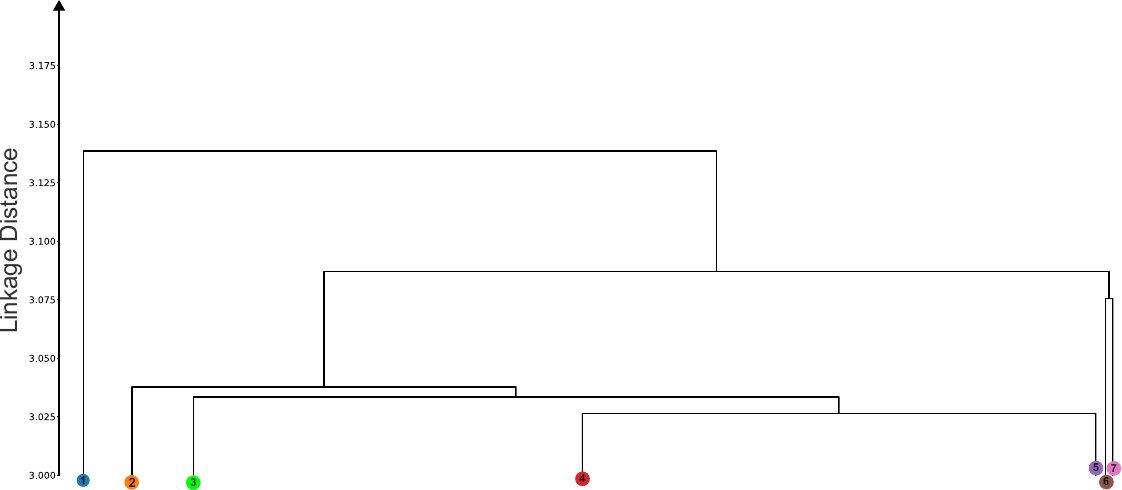}
    \caption{a) Depiction of clustering of barrier configurations using the dendrogram with the linkage distance cutoff at 3.0 \AA.}
    \label{fig:figure_2_clusters}
\end{figure}
Given that all our input features are based on interatomic distances in the LTO crystal, the distances between the clusters depicted in Fig.~\ref{fig:figure_2_clusters} represent distances in the corresponding feature space. We therefore use a cutoff of 3.0~\AA{}, corresponding to the minimum Ti~--~Ti distance between two distinct polaron sites, to identify clusters of polaron configurations. This resulted in a total of 7 different clusters as depicted in Fig.~\ref{fig:figure_2_clusters}. Of these clusters, cluster 4 has with 2016 elements the highest number of hopping barriers, while cluster 6 with 18 configurations is the smallest, as depicted in Fig.~S1. 

To ensure our training set's diversity, we next randomly select configurations from each of these clusters.
In total, we selected 161 polaron barrier configurations out of a total of 3649 configurations for training and testing, and for DFT calculation. 
However, 24 polaron barrier configurations were excluded due to problems with the DFT+U barrier estimation, such as inconsistent occupation matrices or a failure to localise polarons in the first place. 
This led to a final training and test dataset of 137 polaron barrier configurations. 
Fig.~S2 shows the bar plot representing the number of barrier configurations calculated using the DFT. Moreover, it also shows a balanced sampling of configurations for training and testing out of the full sample space. Cluster 4 has the highest number (63), and no configurations were considered from Clusters 6 and 7, because the configurations of these clusters proved unstable. For most of these configurations, neither initial nor final polaron localization converged in our DFT calculations, while for others the migrating polaron underwent a physically unlikely spin flip. 
Note that each of the such excluded polaron hops is then also excluded from the subsequent kinetic model described below.

To gain a deeper understanding of how similar barriers configuration are clustered in a cluster, we plotted the DFT calculated minimum, maximum, and average polaron hopping barrier for each cluster in Fig.~S3 of the SI.
We find that cluster 4 has the overall lowest minimum barrier (0.087 eV), while cluster 2 contains the maximum barrier (0.365 eV) configuration. Clusters 3 and 4 have a wider distribution of barriers compared to the other clusters.  This shows that perhaps we might need to decrease the linkage distance cutoff to obtain more distinct clusters to capture more diverse configurations. However, this will also increase the number of clusters and  DFT barriers calculations to train the ML-model. Indeed, making the process computationally more expensive and cumbersome, hence, we stick to the current clustering. However, the diversity offered by our current approach seems to suffice for our purposes, as described below.

\subsection{Machine Learning Model}
For the regression task, we tested three different ML-algorithms; support vector regression (SVR), Gaussian process regression (GPR), and gradient boosted decision trees (GBDT) in order to find the optimal algorithm. For SVM, we chose three different built-in kernels available in the \texttt{scikitlearn} python library;\cite{scikit-learn} radial basis function (RBF), linear, and polynomial --- see supporting information. For GPR, we used RBF and a combination of constant kernel (CK), RBF, and white kernel (WK), as also outlined in the supporting information. Unfortunately, both approaches, SVM and GPR, performed quite poorly on our regression task, i.e.,~the prediction of polaron hopping barriers. Alternatively, we used the GBDT algorithm implemented in the \texttt{XGBoost} python library \cite{chen2016xgboost}. GBDT is an ensemble learning method that combines multiple weak learners (decision trees) to create a strong predictive model. It trains the model in a sequential manner, where each successive tree is built to correct the errors made by the previous trees. More precisely, it subsequently fits the decision trees to the negative 
gradient of the loss function and hence iteratively minimizes the prediction error. XGBoost can perform both regression and classification tasks, and for regression, we set the \texttt{objective = "reg:absoluteerror"} which minimizes the loss function defined as: 
\begin{equation}
    L(y,\hat y(x)) =  \frac{1}{N} \sum_{i=1}^{N} \left| y_i - \hat{y}_i\right|
\end{equation}
Here, $y_i$ is the true value and $\hat{y}_i$ is the predicted value for the $i^{th}$ sample. 

In order to train an XGBoost model, we first split the data set into training and test sets. As a first step, we trained our model by utilizing the cluster information and performed a train test split by drawing at least one data point from each of the clusters; however, the model's performance for the test set was, at best, moderate. In contrast to our previous work, splitting the data according to clusters did not improve model performance. 
This could be due to the broader distribution of the hopping barriers in each cluster, cf.~barrier distribution shown in Figure S4. Alternatively, to impartially split the train test sets, we first discretized the barriers into 12 equal bins by using the \emph{pandas} library's \emph{qcut} module. Furthermore, we performed a stratified split based on the bin, by \emph{sklearn} library. This led us to train model without any unbalanced barrier domains.
This resulted in a training set of 123 configurations and an independent test set of 14 configurations. To assess the robustness of the model with respect to the choice of training and validation configurations, we performed 10 independent random splits of the 123 training configurations, with 110 configurations used for training and 13 for validation in each split. An \texttt{XGBoost} model was trained independently for each split, and its performance was evaluated using the validation RMSE. Across the 10 splits, the mean validation RMSE was 0.051 eV with a standard deviation of 0.004 eV. The relatively low mean RMSE and small variation across the different splits indicate that the model provides consistent predictions of the polaron hopping barriers and is relatively insensitive to the specific training/validation partition. Given that, we trained our final model with the 123 training set by optimizing the hyperparameters using the \texttt{GridSearchCV} function with a five-fold cross-validation. 

\begin{figure}[htbp]
\centering
\includegraphics[scale=0.8]{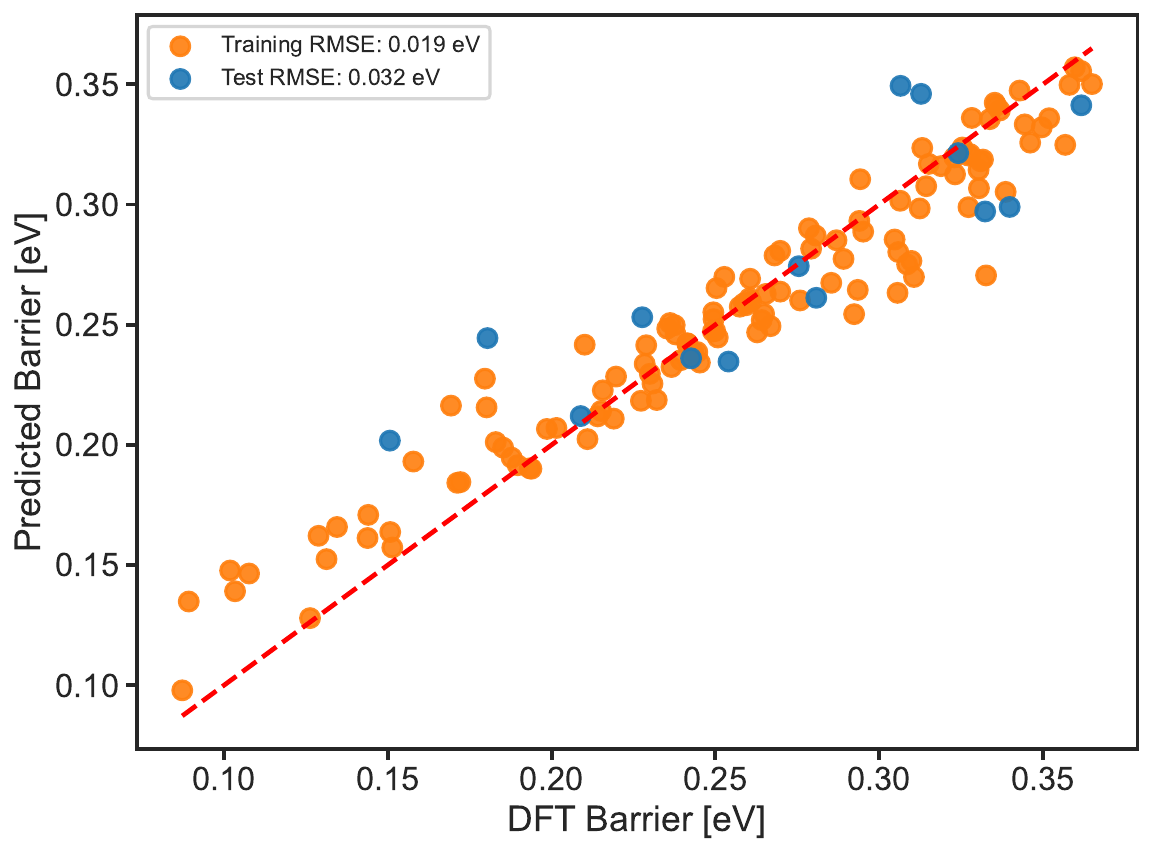}
\caption{Correlation plot for training and test dataset using the XGBoost. The RMSE for training and test dataset is 0.019 eV and 0.032 eV, respectively.}
\label{fig:xgb_test_train_corr}
\end{figure}

Thus, the final set of optimized hyperparameters are: \texttt{n\_estimators = 450}, \texttt{max\_depth = 4}, \texttt{learning\_rate = 0.02}, \texttt{subsample = 0.4},  \texttt{colsample\_bytree = 0.4}, \texttt{reg\_alpha = 0.05}, and \texttt{reg\_lambda = 0.4}. Note that a moderate number of decision trees (\texttt{n\_estimators}) with moderate depth (\texttt{max\_depth}) are required to capture the complex relationship between the structural features and the polaron barriers. A small learning rate ensures gradual learning, while \texttt{subsample} and \texttt{colsample\_bytree} aid in the consideration of a random and diverse subset of configurations and features, again helping to reduce overfitting. Finally, we also set the regularization parameters (\texttt{reg\_alpha} and \texttt{reg\_lambda}) to further reduce overfitting. The performance of the final model on the training and completely unseen test set is shown in Fig.~\ref{fig:xgb_test_train_corr}, with an RMSE of 0.019 eV and 0.032 eV, respectively. This indicates that the model is able to accurately predict polaron hopping barriers based on the selected structural features.

Note that, for both the train and the test sets the low barrier regime is slightly overestimated, which is likely due of the broader range of the hopping barriers. Luckily, the polaron hopping in LTO is dominated by the barriers around 0.25 eV, so such an overestimation for a few configurations is acceptable.

\begin{figure}[htbp]
\centering
\begin{overpic}[width=0.49\textwidth]{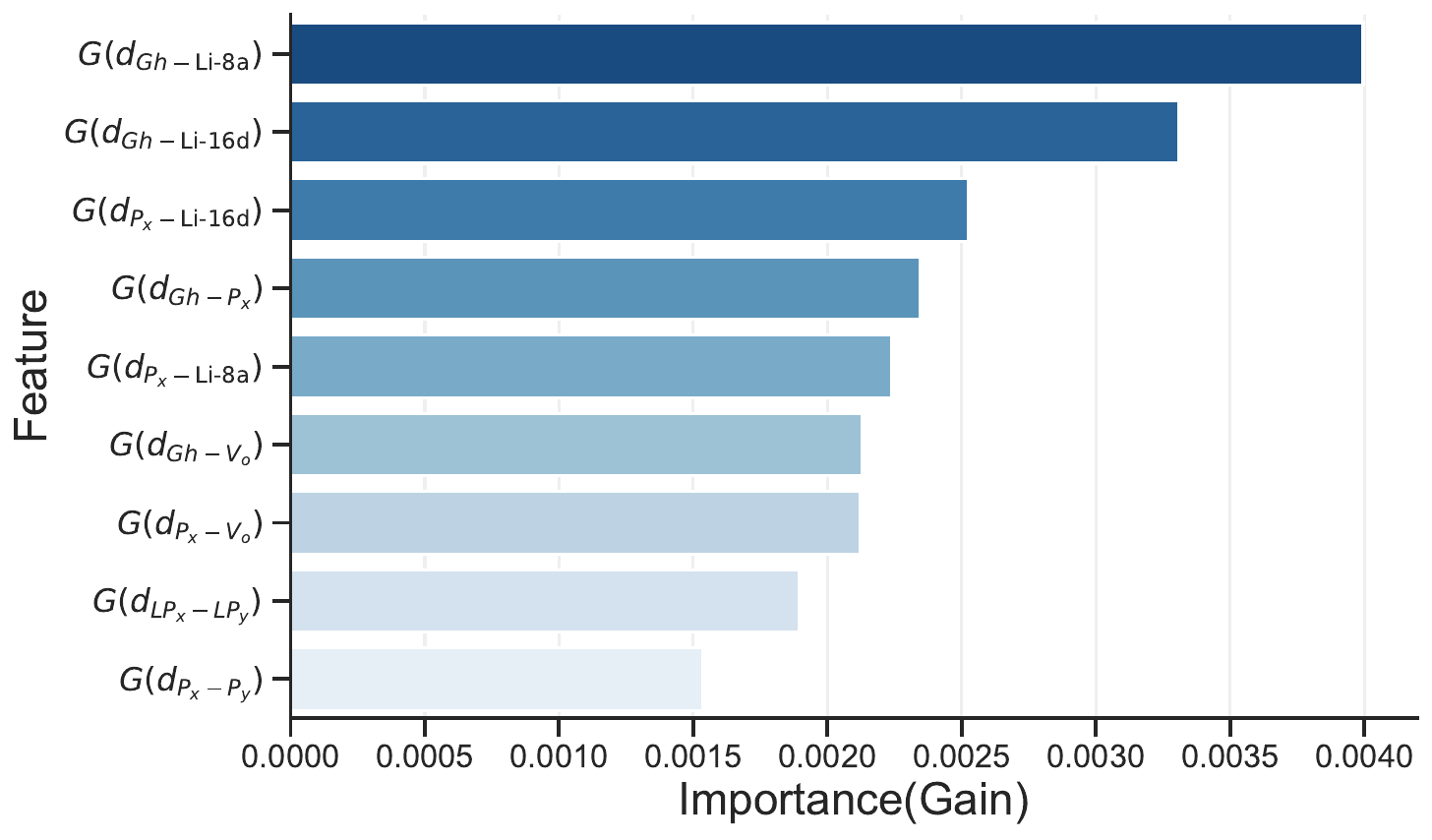}
  \put(10,58){\textbf{(a)}} 
\end{overpic}
\begin{overpic}[width=0.49\textwidth]{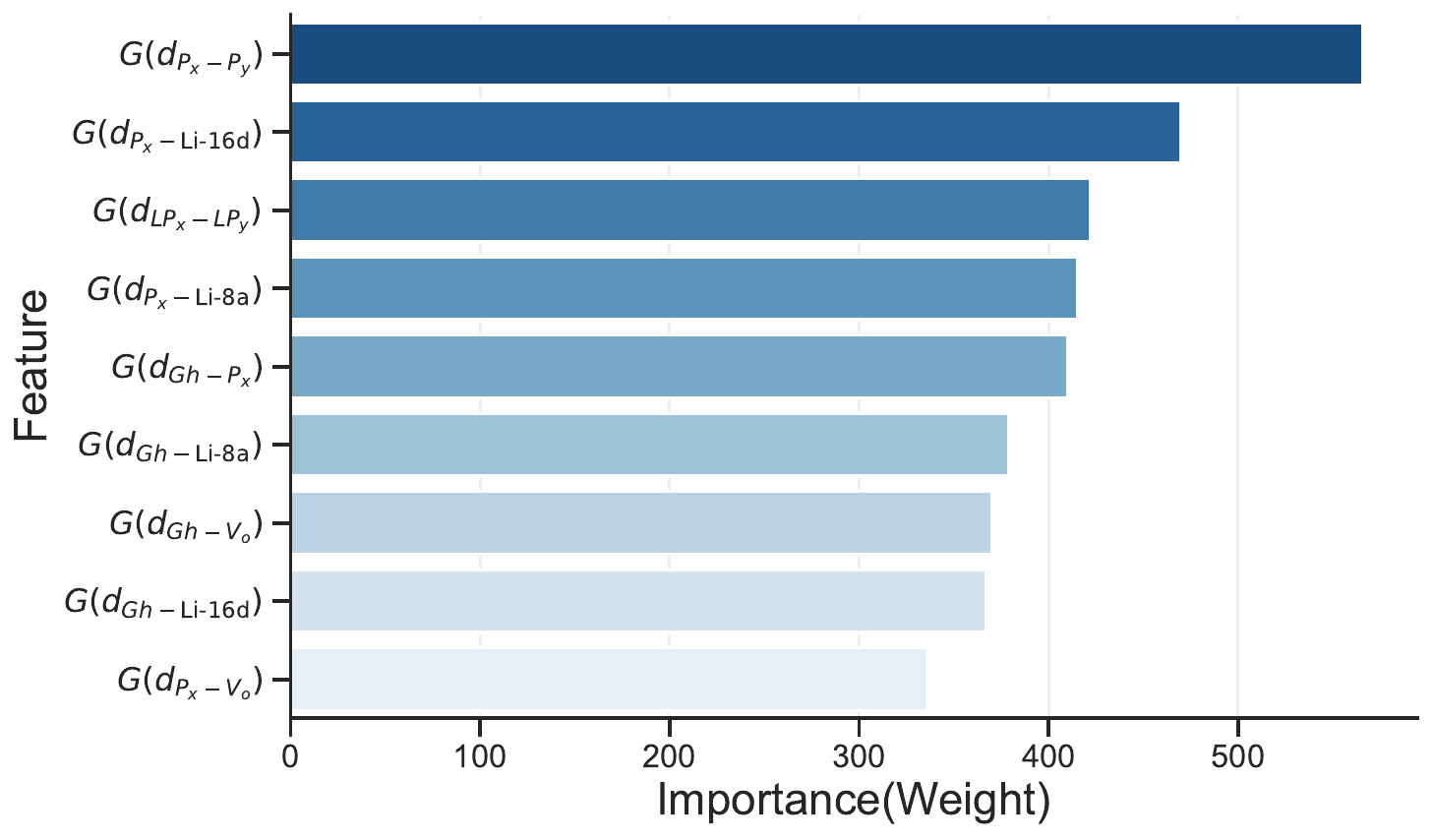}
  \put(10,58){\textbf{(b)}}
\end{overpic}                   
\caption{Feature importance plot obtained from the XGBoost model. (a) Gain, and (b) Weight.}
\label{fig:feature_importance}
\end{figure}

Furthermore, to obtain deeper insights into the selected features, we plot the importance of features extracted from the XGBoost model, c.f.~Fig.~\ref{fig:feature_importance}. The \texttt{gain} based feature importance (Fig.~\ref{fig:feature_importance}a), which reflects the improvement in the model's accuracy due to a particular feature during a split. It indicates that the distance between the ghost atom and the Li-8a ion is the most important characteristic, followed by the ghost atom and the Li-16d ion distance. This indicates the strong correlation between the polaron hopping barriers and Li-ions.

The \texttt{weight} based feature importance depicted in Fig.~\ref{fig:feature_importance}b) reflects the number of times a feature is used to split the data across all trees in the model. It indicates that the polaron-polaron distance ($G(d_{P_x-P_y})$) is the characteristic most considered to make a split, subsequently the distance between the polarons and the Li-16d ions ($G(d_{P_x-\text{Li-16d}})$). Furthermore, the vertical distance between the polarons had its important during the splits.

\section{Results}
\subsection{Hopping in Ti-Layer}
\begin{figure}[htbp]
\centering  
\begin{overpic}[width=0.38\textwidth, trim=10pt 5pt 110pt 5pt,  
  clip]{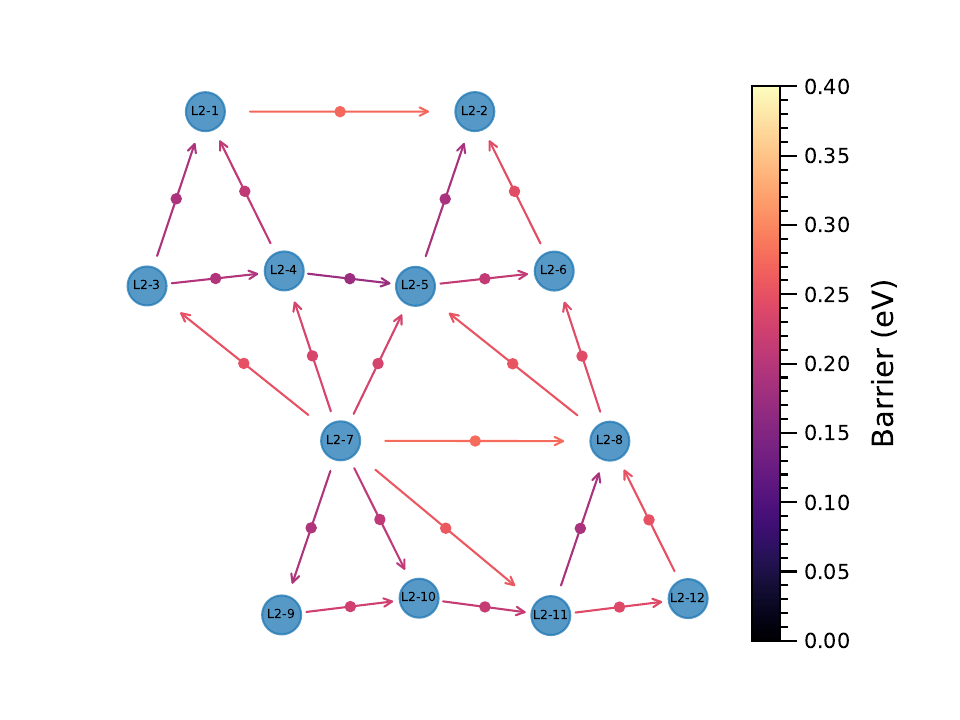}
  \put(-2,83){\textbf{(a)}} 
  \put(30,-2.5){Static Pol: L4-1}
  \put(6,5){[001]}
\end{overpic}
\begin{overpic}[width=0.49\textwidth, trim=10pt 5pt 10pt 5pt,  
  clip]{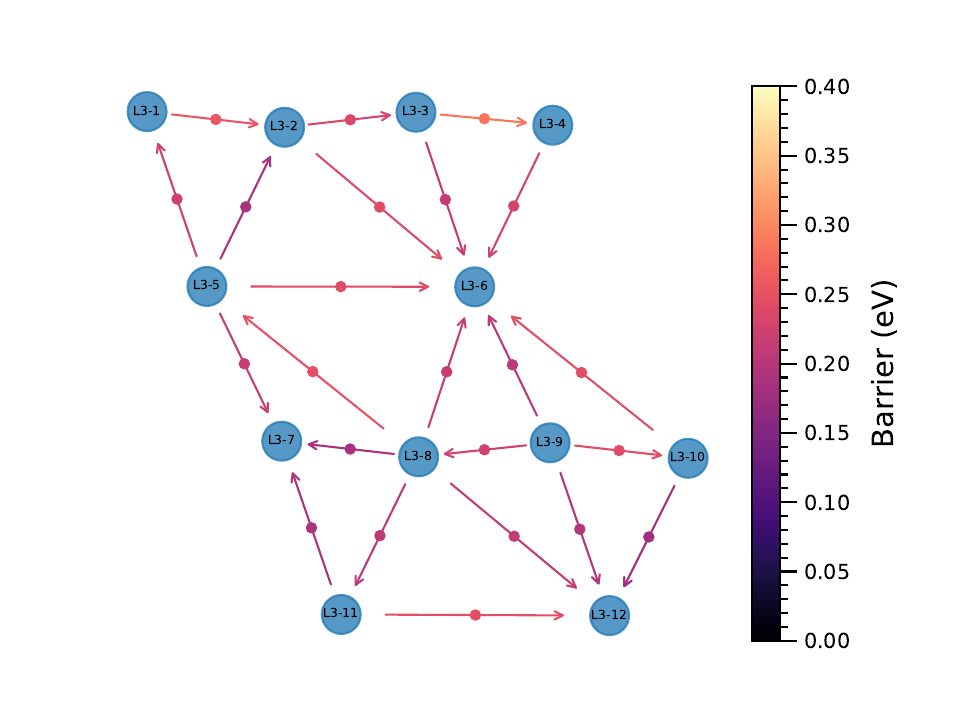}
  \put(-7,65){\textbf{(b)}}
  \put(30,-2.5){Static Pol: L5-1}
  \put(12,5){[001]}
\end{overpic}                   
\caption{Schematic representing Ti-atoms and corredspong intra Ti-layer polaron hopping barrier projected on a 2D plot, a) for Ti-layer L2 and b) for Ti-layer L3.}
\label{fig:intra_layer_hopping_L2_L3}
\end{figure}

Next, in order to get more insight about the polaron migration barrier in the LTO lattice, we plotted the predicted polaron hopping barrier among Ti-atoms in the intra Ti layer. This involved projecting each Ti layer from the LTO crystal to a 2-dimensional (2D) structure and the corresponding hopping barriers to the neighboring site. The static polaron for these barrier configurations was considered at a distance of two Ti-layers -- For L1, L2, L3, L4 and L5 layers, the static polaron was at L3-1, L4-1, L5-1, L1-1, and L2-1, respectively. A schematic of such projected atomic positions for the L2 and L3 layers and the corresponding barriers is shown in the Fig.~\ref{fig:intra_layer_hopping_L2_L3}. The colored arrows represent the hopping direction, the color represents the barrier strength and the corresponding values are shown in the color bar.

It is evident from the figure that for both layers the hopping barrier ranges from 0.173 to 0.285 eV, with the lowest \texttt{L2-4->L2-5} in \texttt{L2} and the highest barrier for \texttt{L3-3->L3-4} in \texttt{L3}. We can see from the figure that the farther the initial and final site is, the higher the barrier, and the closer the configurations, the lower the barrier.
 Similarly, we plotted the schematics for \texttt{the L1}, \texttt{L4} and \texttt{L5} layers, shown in Fig.~S7. 
For layers \texttt{L1} and \texttt{L4}, they have a similar and high barrier of $\sim$ 0.30 eV. However, interestingly, the Ti-layer \texttt{L5} has shown a much wider range of hopping barriers, ranging from 0.14 to 0.24 eV; see Fig.~S7~b. 
The barrier configurations \texttt{L5-1->L5-2}, \texttt{L5-2->L5-3} and \texttt{L5-3->L5-4} have barriers of 0.20, 0.14, and 0.15 eV, respectively. This could be because these configurations are the closest to the Oxygen vacancy, which highlights the importance of including the feature $G(d_{P_x-V_o})$.  
A similar analysis for inter-layer vertical polaron hopping is depicted in Fig.~S8~a). The corresponding hopping barriers are shown in Fig.~S8 (b-e). 
Interestingly, only a few configurations that are closer in distance have a lower migration barrier. There is only one configuration \texttt{L3-5|L4-4->L3-10} with a migration barrier of 0.20 eV, while the rest of the configurations have a barrier greater than 0.20 eV.  This higher polaron migration barrier along the z-direction hints at the preferred polaron migration in parallel to the xy-plane of the simulation cell.   

\subsection{Polaron Mobility}
\begin{figure}
    \centering
    \includegraphics[scale=0.7]{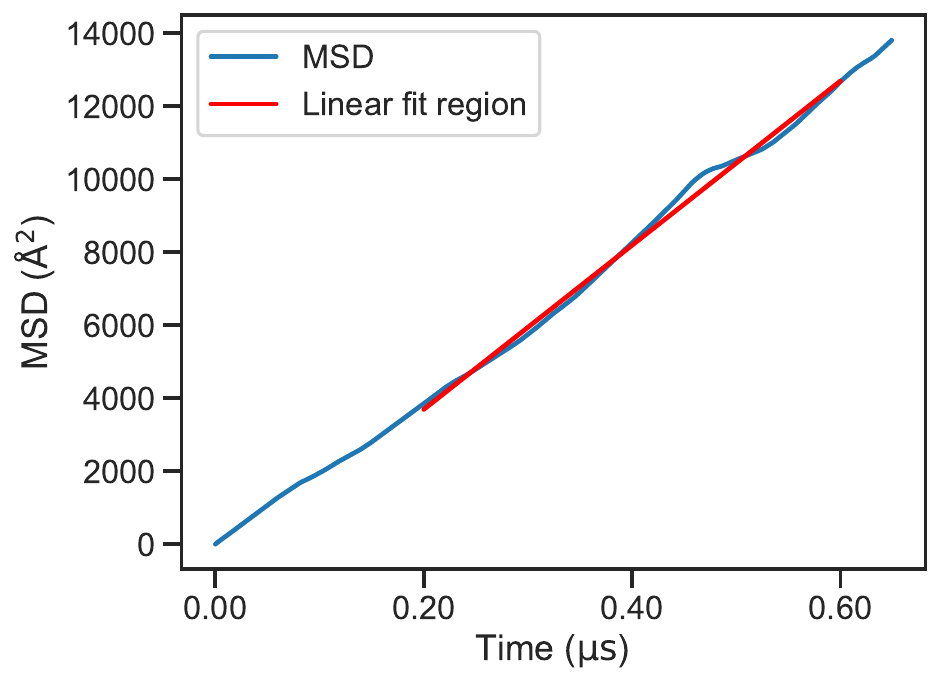}
    \caption{MSD vs time plot for the polarons in LTO at 300 K. The diffusion coefficient was calculated by fitting the linear region of the plot.}
    \label{fig:msd_plot}
\end{figure}

With the trained ML-model, we can predict the hopping barriers for all of the 3649 configurations. These, we now utilize to calculate some physical observables of interest, such as the diffusion coefficient, and the remaining properties can be derived from it. To calculate the diffusion coefficient, we perform kinetic Monte Carlo (kMC) simulations using an in-house code based on the description of Van der Ven \emph{et al.} \cite{van2001first}. The kMC simulations were performed for a 2$\times$2$\times$1 supercell with periodic boundary conditions. The hopping rate for each jump from site $i$ to site $j$ is calculated using the Arrhenius rate equation:\cite{devreese1996encyclopedia}
\begin{equation}
    k_{ij} = \nu \exp\left(-\frac{E_{a,ij}}{k_B T}\right)
\end{equation}
Where, $\nu$ is the attempt frequency defined as $ \nu = \frac{k_B T}{h}$ and $E_{a,ij}$ is the activation energy for the jump from site $i$ to site $j$ obtained from the ML model. We ran the kMC simulation for $10^{-6}$ sec at a temperature of 300 K. To calculate the diffusion coefficient (D) we simply employed the Einstein relation:
\begin{equation}
    D = \frac{1}{\text{2d}} \lim_{\tau \to \infty} \frac{d}{d\tau} \text{MSD}
\end{equation}
where, d is the dimensionality of the system (d=3 for 3D diffusion) and MSD is the mean-squared displacement (MSD) for the polarons. For only two polarons the MSD calculated with a single time origin suffered from poor statistics. Thus, we used the multiple time origin method to calculate the MSD, defined as\cite{pranami2015estimating}:
\begin{equation}
    \text{MSD}(\tau) = \frac{1}{N_{\tau}} \sum_{i=1}^{N_{\tau}} \left( \frac{1}{M_p} \sum_{j=1}^{2} |\mathbf{r}_i(t_j + \tau) - \mathbf{r}_i(t_j)|^2 \right)
\end{equation}
Where, $N_{\tau}$ is the number of time intervals of length $\tau$ and $M_p$ is the number of polarons in the simulation cell. The calculated MSD vs time plot has been shown in Fig.~\ref{fig:msd_plot}. Initially, the polaron dynamics follows the Fickian diffusion with a slight saturation ($t\sim 0.50 \mu$s), followed again the Fickian diffusion. This is evident from the linear region of the plot. Moreover, the slight saturation highlights the polaron trapping between the hopping sites. 
Given that, we calculated the  diffusion coefficient by fitting the linear region of the MSD vs time plot. 
The diffusion coefficient was found to be $D = 3.751 \times 10^{-7}$ cm$^2$/s.

From this, the mobility ($\mu$) was calculated using: 
$\mu = q\times D/(K_{B}T)$
to be $1.450\times$10$^{-5}$ cm$^2$/Vs. Moreover, we also calculated the conductivity ($\sigma$ = $nq\mu$), where $n$ is the charge carrier concentration. For two plarons in the 2$\times$2$\times$1 supercell, the charge carrier concentration was calculated to be $n = 1.11 \times 10^{21}$ cm$^{-3}$ which is right in the experimentally plausible range. \cite{kick2020mobile} 
This resulted in a conductivity of $\sigma = 2.576 \times 10^{-3}$ S/cm. 
Our calculated conductivity is, thus, in good agreement with experimental results for reduced LTO, which in the literature range from $10^{-5}$ S/cm\cite{wolfenstine2006electrical} to $10^{-2}$ S/cm\cite{scharner1999evidence}, yet with the latter value being achieved with additional lithiation and therefore an increase of polaron density. Our results have to be understood as idealised values without grain boundaries as would be prevalent in the measured LTO powders. While a similar approach to ours could be applied to grain boundaries, this lies outside the scope of the current study.

Finally, we also computed the temperature variation of the diffusion coefficient, as shown in Fig.~S9 (a). Unsurprisingly, given the activated nature of our dynamics, we find the diffusion coefficient to increase increase in temperature, while decrease as the temperature decreases; following the Arrhenius behaviour. The fitted activation energy from the Arrhenius plot was calculated to 0.22 eV. This value reflects the influence of the DFT-derived energy landscape on the macroscopic diffusion process and represents an effective activation energy describing its temperature dependence, rather than an individual hopping barrier.

\begin{table}[h!]
\centering
\caption{Diffusion coefficients~(D), Mobility~($\mu$), and conductivity~($\sigma$) computed with differently approximated hopping barriers, comparing ML barriers to the minimum, median, and maximum of the DFT computed barrier distribution.}
\begin{tabular}{lcccc}
\toprule
S.No. & Barrier (eV) & D~(cm$^2$/s) & $\mu$~(cm$^2$/Vs) & $m\sigma$~(S/cm) \\
\midrule
1 & ML           & 3.751$\times10^{-7}$ & 1.450$\times10^{-5}$ & 2.58 \\
2 & Min.~(0.125) & 2.234$\times10^{-5}$ & 8.640$\times10^{-4}$ & 153  \\
3 & Med.~(0.247) & 1.308$\times10^{-7}$ & 5.057$\times10^{-6}$ & 0.90 \\
4 & Max.~(0.357) & 4.083$\times10^{-9}$ & 1.579$\times10^{-7}$ & 0.03 \\

\bottomrule
\end{tabular}
\label{tab:kmc_data_diff_barr}
\end{table}
 
Next, to assess the importance of our machine-learned, configuration-dependent polaron hopping barriers in the kMC simulations, we performed a series of controlled simulations by fixing the barrier to the minimum (0.087 eV), median (0.261 eV) and maximum (0.365 eV) values of the distribution of DFT-calculated barriers. The simulated results for temperature 300~K are given in table~\ref{tab:kmc_data_diff_barr}, and the temperature variation at different fixed barriers is shown in Figures~S9 (b-d). From table~\ref{tab:kmc_data_diff_barr}, we can see that there is a strong dependence of the diffusion coefficient on the hopping barrier which is consistent with the Arrhenius character of the hopping rate. Increasing the barrier from the minimum to the maximum value decreases the diffusion coefficient by four orders of magnitude. Similarly, the mobility and the conductivity follow the same trend. This demonstrates the sensitivity of long-range polaron transport on the polaron hopping barriers and highlights the importance of barrier prediction of each hopping configuration. In addition to this, the median barrier results in a conductivity of 0.90~$m$~S/cm, while the ML barrier for each configuration results in a conductivity of 2.58~$m$~S/cm. This highlights that conductivity cannot necessarily be inferred simply from the statistical median of the barrier distribution.
The barrier values obtained from the kMC simulations are slightly higher than the actual given values. This behavior demonstrates that the diffusion process is sensitive to the underlying barrier landscape. As a last analysis step, we plot the normalized count of the number of times each polaron site was visited during the kMC simulation at 300~K for the actual ML barriers and the fixed barriers. The bar plots for the site counts are shown in Figure~S10~(a-d). For actual ML barriers among the migration sites, the polaron has visited the least on site \texttt{L1-1}, and the most on sites \texttt{L5-3}, \texttt{L5-2}, and \texttt{L5-4}. These frequent visits of polaron are justified by the migration barriers between these sites, c.f. Fig.~S8. However, for fixed barriers, the polarons explore the localization sites almost equally and would lead to different macroscopic transport properties.  This further supports the validity of our approach, and we can conclude that the conductivity in defected-LTO mainly arises due to the small polarons. 

\section{Conclusion} 
In conclusion, in this study we have investigated the small polaron dynamics in the battery material LTO using a combination of DFT calculations, ML, and kMC simulations. Taking into account the pitfalls of small data learning, we first clustered all the possible polaron migration configurations using the agglomerative clustering method and selected a diverse set of 161 configurations for DFT calculations. Of the 161 configurations, only 137 configurations were considered for the ML model train and test purpose. Specifically, we trained our ML model using 123 data points and tested it with 14 points. Among the different ML models, XGBoost outperformed in the regression task, with an RMSE of 19 meV and 32 meV for the test and training datasets, respectively.

An analysis of the \emph{gain}-based feature-importances revealed that the migration barrier of the polaron is most strongly determined by the interaction between the ghost-atom (mid-point of the reaction path) and the Li-8a ($G(d_{Gh-\text{Li-8a}})$) and Li-16d ($G(d_{Gh-\text{Li-16d}})$) ions.  This highlights a correlation between the polaron migration barrier and the Li-ions. In addition, a weight-based feature analysis revealed that the presence of other polarons in the vicinity of the migrating polaron ($G(d_{P_x-P_y})$) also plays a significant role in determining the migration barrier. 

The trained XGBoost model was then utilized to predict migration barriers for all 3649 configurations. The predicted polaron migration barriers in LTO range from 0.135 to 0.357 eV with a mean of 0.25 eV. A structural analysis based on the predicted barriers showed that Ti-layers \texttt{L1} and \texttt{L2} tend to have high barriers of $\sim 0.30$ eV, while \texttt{L5} shows the lowest. This is due to the influence of the nearby oxygen vacancy.
Compared to in-layer migration, polaron movement along the z-direction has most configurations with barriers greater than 0.20 eV. This implicitly points to the directional dependence of the polaron migration.

Finally, physical carrier dynamic observables were obtained from kMC simulations performed with barriers predicted by the ML model. The calculated diffusion coefficient for LTO was found to be $3.751\times10^{-7} cm^{2}/s$. A carrier concentration of $n = 1.11\times10^{21} cm^{-3}$ for two polarons led to a conductivity at 300K temperature of $\sigma = 2.576 \times 10^{-3}$ S/cm, which is in good agreement with the experimental study carried out on reduced LTO doped with Ta. The mean barrier calculated by the Arrhenius analysis was found to be 0.22 eV, which is slightly lower than the mean of all predicted ML barriers. This highlights that the diffusion process is sensitive to the barrier landscape. 
To verify the importance of our ML approach versus more simple ways of approximating barriers, we also computed the diffusion coefficients of the kMC simulations with barriers for all sites fixed to the minimum, median, and maximum of the DFT barrier distribution. These obviously showed a strong dependence on the choice of fixed barrier, with even the median barrier of the DFT distribution yielding conductivities off by almost a factor of 3. Overall, this highlights the important and complementary role of ML in materials modeling, extending our reach beyond the prediction of static properties to the efficient exploration of thermally activated polaron hopping dynamics. 

\section{Data Availability}
The datasets generated and analysed during the current study, as well as all trained ML models are available in the Zenodo repository, DOI: https://doi.org/10.5281/zenodo.24157340. 

\section{Acknowledgements} 
H.O.~recognizes support from the Deutsche Forschungsgemeinschaft (DFG, German Research Foundation) within the CRC 1585 MultiTrans. Furthermore, H.O.~and T.S.~gratefully acknowledge funding from the the Bavarian Ministry of Science and Art through the ''Solar Technologies go Hybrid'' consortium and support from the DFG IRTG 2818 OPTEXC. Computations were carried out on the HPC systems Festus (DFG grant number 523317330) of the University of Bayreuth and Fritz (NHR@FAU grant number b243cb, DFG grant number 440719683) of the Erlangen National High Performance Computing Center (NHR@FAU) of the Friedrich-Alexander-Universita\"at Erlangen-Nu\"urnberg (FAU).

\section{Competing Interests}
The Authors declare no Competing Financial or Non-Financial Interests.

\section{Author Contributions}
\textbf{Tavinder Singh}: data curation (lead); formal analysis (lead); investigation (lead); software (lead); visualization (lead); writing -- original draft preparation (lead); writing -- review \& editing (supporting)
\textbf{Harald Oberhofer}: conceptualization (lead); formal analysis (supporting); funding acquisition (lead); investigation (supporting); resources (lead); supervision (lead); writing -- original draft preparation (supporting); writing -- review \& editing (lead)

\bibliography{bibliography}

\includepdf[pages=-]{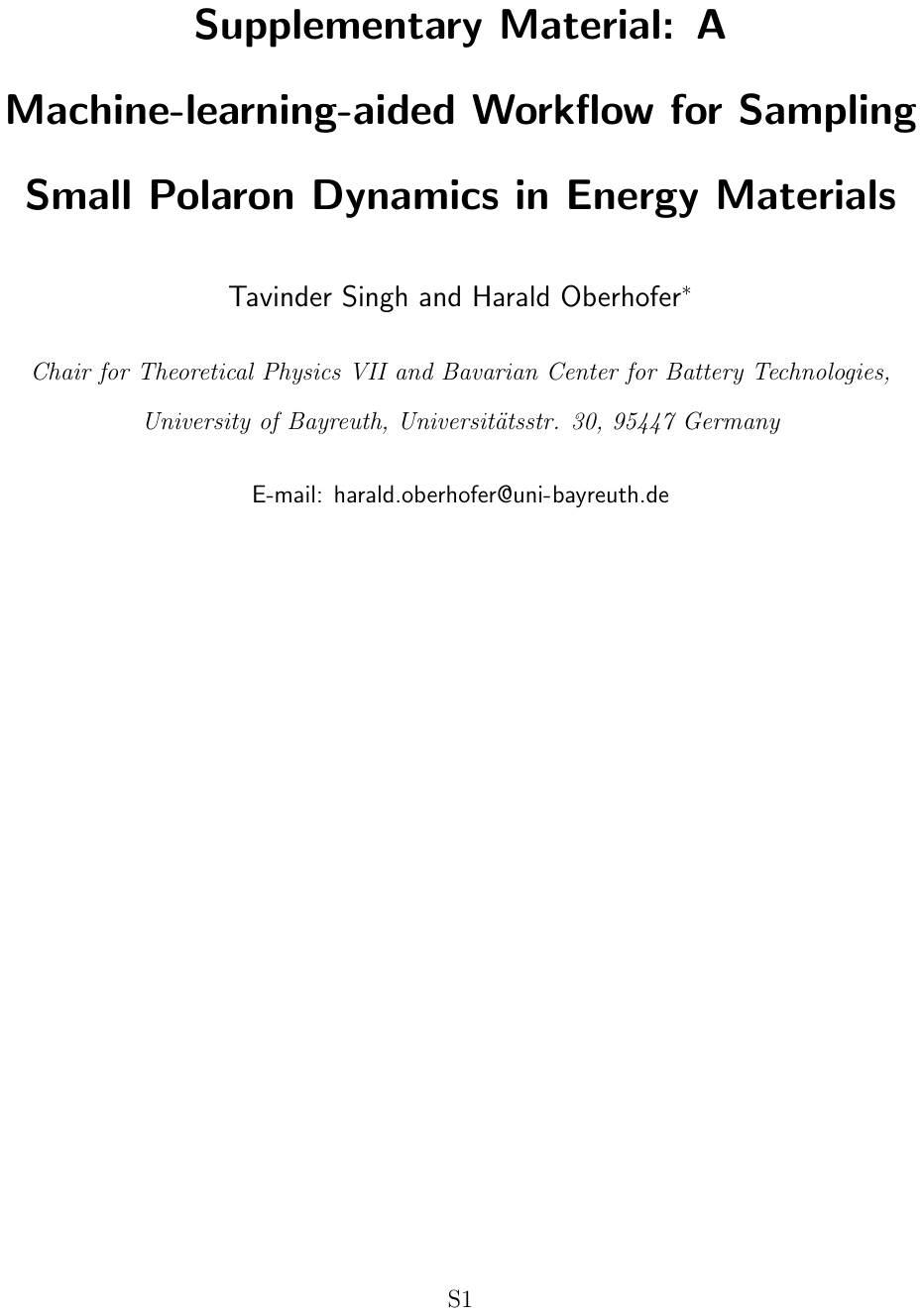}
\end{document}